\documentclass[useAMS,usenatbib]{mnras}

\usepackage{psfig,epsfig,supertabular,wrapfig,placeins}
\usepackage{graphicx,amsmath,amssymb,subfigure,multirow,units}
\usepackage{marvosym,breqn}
\usepackage{graphics,multirow}
\usepackage{amssymb}
\usepackage{amsmath}
\usepackage{euscript}
\usepackage{setspace}
\usepackage{fancyhdr}
\usepackage{afterpage,rotating,url}
\usepackage{lscape,times}
\usepackage{blindtext}

\def\apj{ApJ}
\def\mnras{MNRAS}

\def\araa{ARA\&A}                
\def\aap{A\&A}                   
\def\apjs{ApJS}                  
\def\pasp{PASP}                  
\def\apjl{ApJ}                   

\def\nat{Nature}

\newcommand{\gtsim}{\mbox{{\raisebox{-0.4ex}{$\stackrel{>}{{\scriptstyle\sim}}$}
}}}

\newcommand{\fast}{{\sc Fast}}

\title[SCUBA-2 CLS: SMG redshifts]{A complete distribution of redshifts for sub-millimetre galaxies in the SCUBA-2 Cosmology Legacy Survey UDS field}

\author[D.\,J.\,B.~Smith et al.]
       {D.\,J.\,B. Smith$^{1}$\thanks{E-mail:d.j.b.smith@herts.ac.uk(DS)}, C.C. Hayward$^2$, M.J. Jarvis$^{3,4}$ \&\ C. Simpson$^5$ \\
         $^{1}$Centre for Astrophysics Research, School of Physics, Astronomy and Mathematics, University of Hertfordshire, Hatfield, Herts, AL10 9AB \\
         $^2$Center for Computational Astrophysics, Flatiron Institute, 162 Fifth Avenue, New York, NY 10010, USA \\
         $^3$Oxford Astrophysics, Denys Wilkinson Building, Keble Road, Oxford, OX1 3RH \\
         $^4$Department of Physics, University of the Western Cape, Bellville 7535, South Africa \\
         $^5$Gemini Observatory, Northern Operations Center, 670 North A`\={o}h\={o}ku Place, Hilo, HI 96720-2700, USA
}
\begin{document}

\date{\today}

\pagerange{\pageref{firstpage}--\pageref{lastpage}} \pubyear{2017}

\maketitle

\label{firstpage}

\begin{abstract} 
Sub-milllimetre galaxies (SMGs) are some of the most luminous star-forming galaxies in the Universe, however their properties remain hard to determine due to the difficulty of identifying their optical\slash near-infrared counterparts. One of the key steps to determining the nature of SMGs is measuring a redshift distribution representative of the whole population. We do this by applying statistical techniques to a sample of 761 850\,$\mu$m sources from the SCUBA-2 Cosmology Legacy Survey observations of the UKIDSS Ultra-Deep Survey (UDS) Field. We detect excess galaxies around $> 98.4$ per cent of the 850\,$\mu$m positions in the deep UDS catalogue, giving us the first 850\,$\mu$m selected sample to have virtually complete optical\slash near-infrared redshift information. Under the reasonable assumption that the redshifts of the excess galaxies are representative of the SMGs themselves, we derive a median SMG redshift of $z = 2.05 \pm 0.03$, with 68 per cent of SMGs residing between $1.07 < z < 3.06$. We find an average of $1.52\pm 0.09$ excess $K$-band galaxies within 12 arc sec of an 850\,$\mu$m position, with an average stellar mass of $2.2\pm 0.1 \times 10^{10}$\,M$_\odot$. While the vast majority of excess galaxies are star-forming, $8.0 \pm 2.1$ per cent have passive rest-frame colours, and are therefore unlikely to be detected at sub-millimetre wavelengths even in deep interferometry. We show that brighter SMGs lie at higher redshifts, and use our SMG redshift distribution -- along with the assumption of a universal far-infrared SED -- to estimate that SMGs contribute around 30 per cent of the cosmic star formation rate density between $0.5 < z < 5.0$. 
\end{abstract}

\section{Introduction}
\label{sec:intro}

Sub-millimetre wavelengths are well-suited to studying high redshift galaxies, since redshift effects combine with the steep Rayleigh-Jeans tail of the dust spectral energy distribution resulting in negative $K$-corrections. This ensures that galaxies of a fixed far-infrared luminosity (a proxy for star-formation rate) are approximately equally bright from $z=1$ out to $z=6$ \citep[e.g.][]{blain02}. 

The low resolution of sub-millimetre observations with even the most widely-used cameras -- the Sub-millimetre Common-User Bolometer Array \citep[SCUBA][]{holland99} on the 15m James Clerk Maxwell Telescope, and the Large APEX BOlometer CAmera \citep[LABOCA;][]{siringo09} on the APEX 12m telescope -- and high redshift nature of sub-millimetre galaxies (SMGs), make identifying their counterparts (i.e. the sources responsible for the submillimetre emission) challenging \citep[e.g.][]{ivison07}, particularly in deep optical observations. While the tight far-infrared radio correlation \citep[e.g.][]{helou85,yun01,jarvis10,smith14}, and low sky density of radio sources \citep[e.g.][]{condon12}, allow the unambiguous identification of many SMG counterparts using high-resolution radio observations \citep[e.g.][]{ivison02,ivison05,ivison07}, this is still a challenging undertaking, despite the current availability of sensitive radio data over degree-scale fields. This means that the most fundamental requirement for inferring the properties of SMGs -- their redshift distribution -- is hard to measure \citep[e.g.][]{smail02a}. \citet{chapman05} obtained spectroscopic redshifts to 73 SMGs based on following up 98 out of 104 radio-identified SMGs in a parent sample of 150 $>3\sigma$ SCUBA 850\,$\mu$m sources, while \citet{wardlow11} were able to robustly identify only 72 out of 126 SMGs (an identification rate of 57\,per cent) detected in LABOCA observations of the ECDF-S field, using a combination of radio, 24\,$\mu$m and {\it Spitzer} IRAC data. Recent advances have improved matters, but even the latest methods \citep[e.g. the ``tri-colour" technique put forward by][]{chen16} can identify reliable counterparts to only $69 \pm 16$ per cent of SMGs.  This means that, although SMGs are undoubtedly some of the most rapidly star-forming systems in the Universe \citep[e.g.][]{hayward11,hayward13a}, optical\slash near-infrared studies of their counterparts and therefore redshifts have so far been incomplete. 

At the same time, several studies had attempted to complete the redshift distribution of SMGs using longer wavelengths, for example \citet{koprowski14}, \citet{koprowski16} and \citet{michalowski16}, which used an average SMG template from \citet{michalowski10} to find the redshift giving the best fit to the far-infrared\slash millimetre photometry. Of course, the redshifts produced using long wavelengths are in general less accurate than those produced from optical\slash near-infrared data (although the latter are also highly uncertain at the highest redshifts). There are several reasons for this: for example that far-infrared data are susceptible to the influence of the choice of dust SED, and are subject to errors arising from blended far-infrared photometry (e.g. \citealt{scudder16}, although -- as in the aforementioned studies -- this may be mitigated using deblending techniques such as those from \citealt{merlin15} or \citealt{hurley17}). On the other hand, this method of redshift estimation does benefit from not requiring an optical\slash near-infrared counterpart, is able to obtain individual redshifts for every source, and is also potentially less susceptible to the influence of chance alignments between optical sources, e.g. lensing. 

Given the low-resolution of sub-millimetre data based on single-dish observations, interferometers have an important role to play for studying SMGs. Some of the first results using sensitive data from the Atacama Large Millimetre Array (ALMA) to identify the counterparts to SMGs by \citet{hodge13} 
showed using a combination of radio and mid-IR data that $\sim 80$ per cent of SMGs are identified correctly, but with completeness of just 45 per cent. Interferometry has also shown that many SMGs have multiple counterparts; while \citet{barger12} found that 37.5 per cent of bright SMGs are multiples (albeit with a 1$\sigma$ confidence interval between 17-74 per cent),  \citet{hodge13} found that at least 35 per cent of SMGs are multiples, while \citet{chen13} used
the SMA to determine a lower multiple fraction of $12.5^{+12.1}_{-6.8}$ per cent, and this value is likely to be affected by the sample selection (e.g. the flux limit). Multiplicity is also a key parameter if we are to clarify the role of SMGs within current models of galaxy formation and evolution, however it presents considerable challenges for traditional cross-identification efforts \citep[e.g. using likelihood ratio methods][]{wolstencroft86,sutherland92,smith11,fleuren12}. The properties of SMGs are therefore still under investigation, using both observations \citep[e.g. with large amounts of time allocated to investigate the source counts and multiplicity with the Atacama Large Millimeter Array, ALMA;][]{hodge13,karim13,carniani15,simpson15b,fujimoto16,oteo16,dunlop16}, and simulations \citep[e.g. can theory produce sufficiently star forming galaxies, or are chance alignments required?][]{dave10,hayward13b,cowley15,munoz15}. 

New technology used by the Sub-millimeter Common User Bolometer Array 2 \citep[hereafter SCUBA-2;][]{holland13} has given us unprecedented mapping speed at these wavelengths, meaning that we can now efficiently survey large areas. As a result, the SCUBA-2 Cosmology Legacy Survey \citep[CLS;][]{geach17}, has surveyed $\sim$5\,deg$^2$ to better than mJy depth, increasing the total known sample of 850\,$\mu$m sources by an order of magnitude. The extensive efforts to cross-identify these sources at other wavelengths highlights that they are critical if we are to understand the properties of SMGs, and their high-redshift, dusty nature means that the $K$ band data in the UKIDSS Ultra-Deep Survey \citep{hartley13} are ideal for this purpose. With sensitive ancillary data covering the rest-frame UV, optical and near-infrared (UVOIR) bands, we can also estimate the redshifts and stellar masses of possible counterparts, as well as classify them on the basis of their rest-frame colours. \citet{williams09} identified regions in the rest-frame $(U-V)$ vs $(V-J)$ -- hereafter ``$UVJ$" -- space thought to contain predominantly passive and star-forming galaxy populations, and this method is widely used \citep[e.g.][]{whitaker12,hatch14,leja15,mendel15,rees15,fumagalli16}. 

In this paper we statistically study the optical\slash near-infrared properties of galaxies around the positions of a large sample of sub-millimetre sources taken from the CLS. \citet{wardlow11} and \citet{simpson14} made attempts to statistically identify those SMGs that could not be cross-identified using other means in the ECDF-S field, however the ancillary data in those works were not deep enough to detect counterparts to the full SMG sample (while the former work detected counterparts to 85 per cent of their SMG sample, the latter found only $0.61 \pm 0.07$ counterparts per ALMA-blank SMG). Given the availability of the largest samples of sub-millimetre positions from the CLS, and the deepest $K$-band data from the UDS, we now extend this idea to infer the properties of the whole 850\,$\mu$m-selected sample, complementary to the aforementioned long-wavelength estimates of the complete SMG redshift distribution most recently by \citet{michalowski16}.

In section \ref{sec:data} we introduce the data sets used in our
analysis, and explain how our sample was selected, while section \ref{sec:results} contains our results, which we discuss in section \ref{sec:discussion}, before concluding with section \ref{sec:conclusions}. We adopt a standard
cosmology with $\Omega_M = 0.3$, $\Omega_\Lambda = 0.7$, $H_0 =
71$\,km s$^{-1}$ Mpc$^{-1}$, while all magnitudes and colours are
quoted on the AB system \citep{oke83}. 

\section{Data and Sample Selection}
\label{sec:data}

We base our analysis on the 850\,$\mu$m data from the CLS
observations of the UKIDSS UDS field.  The CLS 850\,$\mu$m data
reduction is described in \citet{geach17}, and the UDS
field is centred on $\alpha = 02^h17^m46^s$,
$\delta = -05^\circ05^\prime15^{\prime\prime}$ (J2000), covering an area of 0.8\,deg$^2$ with a median RMS $\sim$0.88\,mJy beam$^{-1}$. We use the catalogue of SMGs in the CLS-UDS maps from \citet{geach17}, which contains 1,088 SMGs detected at $>3.5\sigma$. 

The UDS data release 8
\citep[][]{hartley13} has a $5\sigma$ depth of $K=24.6$ within
2-arcsec apertures over a field of 0.8\,deg$^2$. Over the central 0.6
degree$^2$ of this field, the DR8 catalogue includes matched
photometry in the $UBVRi^\prime z^\prime JHK$, 3.6\,$\mu$m and
4.5\,$\mu$m bands \citep[measured using the method in][]{simpson12}, 
ideal for estimating galaxy colours. We use the high quality ($\Delta z \slash
(1+z) \sim 0.031$) photometric redshifts from \citet{hartley13}, alongside stellar mass estimates and rest-frame colours estimated using the {\sc Fast} code \citep{kriek09}, which will be fully described in Smith et al. (in preparation). We derive these assuming an initial mass function from \citet{chabrier03}, based on the stellar SEDs from \citet{bc03}, with exponentially declining star formation histories, a range of metallicities ($0.2 < Z/Z_\odot < 2.5$), a range of $0.0 < A_V < 5.0$  \citep[assuming a dust law from][]{calzetti00}, and correcting for extinction in our own galaxy. Rest-frame colours for galaxies are estimated by convolving the best-fit BC03 spectra found by \fast\ with the Bessel $U$- and $V$-band curves \citep{bessell90} and with the Mauna Kea $J$ band filter curve \citep{tokunaga02} in the rest-frame. These values allow us to locate galaxies within the $UVJ$ diagram, in which passive galaxies are those which have $(U-V)_{\mathrm{rest}} > 1.3$, $(V-J)_{\mathrm{rest}} < 1.6$ and $(U-V)_{\mathrm{rest}} > 0.88 \times (V-J)_{\mathrm{rest}} + 0.49$.

We limit the influence of bright stars and UDS image artefacts (e.g. detector cross-talk) by removing all sources which are flagged in the UDS masks, supplied on the UDS website\footnote{http://www.nottingham.ac.uk/astronomy/UDS/}. In addition, we remove stellar sources from our sample by using the Bayesian classifications from \citet{simpson13}. We limit our area to the region where the UDS data include photometric redshifts, and after removing regions masked in the UDS catalogue, the effective area of our survey is 0.527\,deg$^2$ in which we find 761 SMGs from the \citet{geach17} catalogue. We also remove those sources with $\chi^2 > 20$ in either the \citet{hartley13} or {\sc Fast} SED fitting, to limit the influence of bad fits on photometric redshifts, galaxy colours and stellar masses.

\section{Results}
\label{sec:results}

In contrast to other investigations on this subject, we do not attempt to identify individual counterparts to CLS sub-millimetre sources. Instead, we take a statistical approach by measuring the properties of galaxies detected within and immediately around the SCUBA-2 beam, after subtracting off the expected background population, as a function of search radius. This method has the benefit of not requiring any assumptions about the number of $K$-band sources around each individual 850\,$\mu$m position, and in the coming sections we will show that this enables us to generate a complete redshift distribution for 850\,$\mu$m sources based on the UDS optical\slash near-infrared photometric redshifts.  However, the price that we pay for this is that we do not identify the counterparts to any individual 850\,$\mu$m source, and we also suffer from the issue that, because the environments of 850\,$\mu$m sources have been shown to be overdense by a factor of $80\pm 30$ times the background level \citep{simpson15b}, the excess galaxies that we see in the near-infrared may not be responsible for the sub-millimetre emission themselves (see section \ref{subsec:distributions}, below). 

\subsection{Astrometry and initial checks}
\label{subsec:astrom}

\begin{figure}
  \centering
  \subfigure{\includegraphics[width=0.49\textwidth, trim=0cm 0.3cm 3cm 3.5cm, clip=true]{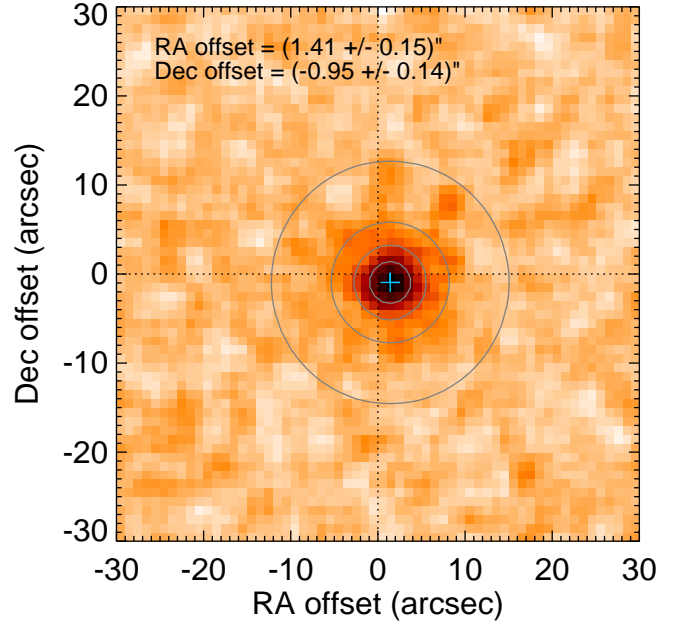}}
  \caption{Image showing a two-dimensional histogram of the differences between the 850\,$\mu$m  positions of the 761 sources in our catalogue and the $K < 24.6$ UDS positions. The dotted lines indicate zero offset in the horizontal and vertical directions, while the blue cross indicates the best fit centroid position, as detailed as in the legend. The grey circles indicate the contours of the best-fit double-Gaussian model used to determine the best fit centroid, with levels corresponding to 0.01, 0.1, 0.2 \&\ 0.5 times the peak value.}
  \label{fig:astrom}
\end{figure}

Before searching around the 850\,$\mu$m source positions for $K$-band galaxies, we must first ensure that the CLS and UDS positions are consistent with each other. We do this by following the method used for sources in the {\it Herschel}-ATLAS survey by \citet{smith11}, which involves creating a 2-dimensional histogram of the difference between the 850\,$\mu$m and UDS positions for every source in each catalogue, which is shown in figure \ref{fig:astrom}. Each bin of the histogram is 1 arc sec in size, and reveals a residual offset in the 850\,$\mu$m astrometry relative to the UDS. To make the best estimate of the magnitude and direction of the offset, we model this distribution using a simple three-component model, consisting of two Gaussians with common centre, plus a constant background, and find the best fitting values for the  centroid. The best-fit offsets derived using this model are $\Delta \alpha = 1.41 \pm 0.15$ arc sec and $\Delta \delta = -0.95 \pm 0.14$ arc sec, giving an effective offset of 1.7 arc sec (i.e. smaller than the pixel size in the 850\,$\mu$m maps) which we apply to the CLS catalogue before continuing. 

\subsection{The fraction of 850\,$\mu$m sources detected in the UDS catalogue}
\label{subsec:q0}

\begin{figure}
  \centering
  \subfigure{\includegraphics[width=0.49\textwidth, trim=0cm 0cm 0cm 0cm, clip=true]{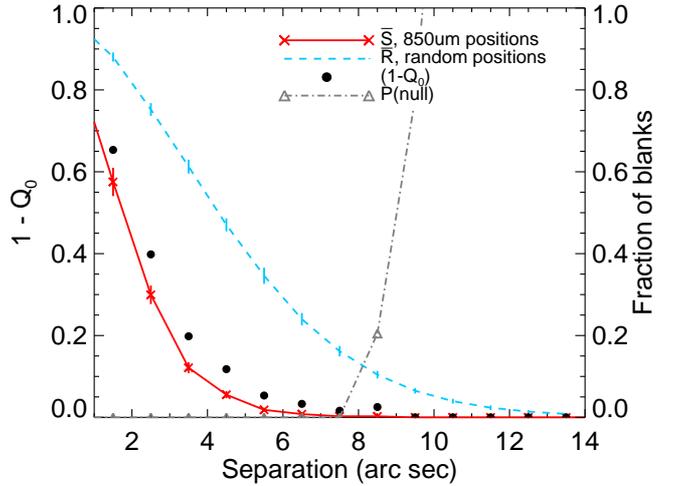}}
  \caption{Illustration of the procedure of \citet{fleuren12} to estimate $(1-Q_0)$, the fraction of 850\,$\mu$m sources which are not detected in the UDS catalogue. The red crosses show the fraction of blanks (i.e. those 850\,$\mu$m positions with no possible counterpart within the search radius), while the light blue dashed line shows the corresponding values based on random positions, both relative to the right-hand vertical axis. The black filled circles represent the values obtained dividing the number of blanks by the randomly generated values, which is our estimate of $(1-Q_{0})$, relative to the left-hand axis. The grey dot-dashed line linking the triangles, indicates the probability of obtaining the number of blanks around the 850\,$\mu$m source positions by chance, and this is $< 1$\,per cent at all separations $\le 10$ arc sec. The values for $Q_0$ and the probability of the null hypothesis are shown in table \ref{table:q0}.}
  \label{fig:q0}
\end{figure}

Next, we determine how many of the 850\,$\mu$m sources have at least one potential counterpart in the UDS catalogue. We do this using the method of \citet[hereafter F12]{fleuren12}, in which this parameter is referred to as $Q_0$. It is difficult to estimate this value directly, due to the the high probability of observing unrelated background sources around the 850\,$\mu$m positions (i.e. the observed number of apparent counterparts within a particular search radius is in fact the sum of the actual number of counterparts and the number of chance associations), and because of the possibility of finding multiple $K$-band galaxies per 850\,$\mu$m source. F12 showed that it is preferable to estimate $(1-Q_0)$, by instead counting the number of `blanks' (i.e.  the number of sources without any possible counterpart) and statistically accounting for the unrelated background population. In the F12 formalism: 

\begin{equation}
	(1 - Q_0) = \frac{\bar{S}}{\bar{R}}, 
	\label{eq:f12}
\end{equation}

\noindent where $\bar{S}$ is the fraction of 850\,$\mu$m sources without a possible counterpart, and we estimate $\bar{R}$ (the fraction of random positions without a possible counterpart) by generating one hundred model catalogues with 761 random positions within the survey area, and returning the number of times zero matches are obtained (accounting for the masked area of our survey). We estimate these values as a function of radius, and the results are shown in figure \ref{fig:q0}, in which the light blue dashed line and red crosses connected by a red solid line show the fraction of blanks around random positions and 850\,$\mu$m positions, respectively. The red crosses (showing the results around the 850\,$\mu$m positions) differ considerably from the results around random positions.  

\begin{table}
\centering
	\caption{Tabulated values of $Q_0$ as a function of maximum separation from the 850\,$\mu$m source positions, with the right-hand column showing the probability of obtaining our results by chance, after accounting for the fraction of sources in the $> 3.5\sigma$ 850\,$\mu$m catalogue which are spurious.}
	\begin{tabular}{c|cc}
		\hline
		Max separation & \multirow{2}{*}{$Q_0$} & \multirow{2}{*}{$P_{\mathrm{null}}$}\\
		 (arc sec) &   \\
		\hline
		1 & 0.103 & 0.000\\
		2 & 0.346 & 0.000 \\
		3 & 0.602 & 0.000\\
		4 & 0.802 & 0.000 \\
		5 & 0.882 & 0.000\\
		6 & 0.947 & 0.000\\
		7 & 0.967 & 0.000 \\
		8 & 0.984 & 0.000  \\
		9 & 0.975 & 0.205 \\
		10 & 1.000 & 0.780  \\
		\hline
	\end{tabular}
	\label{table:q0}
\end{table}

The black filled circles show our best estimate of $(1 - Q_0)$ for CLS 850\,$\mu$m sources, estimated using equation \ref{eq:f12}, and detailed in table \ref{table:q0}. To determine whether seeing an excess $K$-band galaxy around all of the 850\,$\mu$m sources is real (or just due to the large source density in the UDS catalogue), we first estimate the average number of excess non-blanks (i.e. detections) that we observe above what would be expected at random positions, $N_{\mathrm{excess}}$. We do this using equation \ref{eq:nexcess}:

\begin{equation}
N_{\mathrm{excess}} = (\bar{S} - \bar{R})N_{\mathrm{850}} - \sum_{N_{\mathrm{850}}} P(\mathrm{false}),
\label{eq:nexcess}
\end{equation}

\noindent where $N_{\mathrm{850}}$ is the number of 850\,$\mu$m sources in our sample, and $P(\mathrm{false})$ is the probability that each 850\,$\mu$m source is a false positive \citep[as calculated by][and these values indicate a total of $45.6 \pm 1.8$ spurious sources in our sample]{geach17}. We then calculate the null probability of this excess by taking the probability that a position is a blank out to some separation, $\bar{R}$, and multiplying it by itself $N_{\mathrm{excess}}$ times, giving a null probability of $P_{\mathrm{null}} = \bar{R}^{N_{\mathrm{excess}}}$. The results are shown by the grey dot-dashed line in figure \ref{fig:q0}, and tabulated for separations $\le 10$ arc sec in table \ref{table:q0}, which suggests that we can be confident that we detect a significant galaxy excess around $>98.4 \pm 0.9$ per cent of the 850\,$\mu$m sample. This value is a little surprising, since the `false-positive' probabilities imply that $Q_0$ should be no more than 94 per cent, and the uncertainty on $Q_0$ is insufficient to fully account for the discrepancy. If the false-positive probabilities are unduly conservative, this is relevant to other studies as well, for example \citet{michalowski16}, who found that the faintest sub-millimetre sources have a lower ID fraction, and attributed this partly due to the increased fraction of spurious sources at fainter flux limits, and partly due to fainter sub-millimetre sources also being fainter in the radio maps (or other ancillary data) and thus harder to cross-identify. Forthcoming observations with ALMA will hold the key to determining how many of the sources in this catalogue are truly real.

\subsection{The $K < 24.6$ galaxy population around 850\,$\mu$m sources}
\label{subsec:multiplicity}

\begin{figure}
  \centering
  \subfigure{\includegraphics[width=0.49\textwidth, trim=0cm 0cm 0cm 0cm, clip=true]{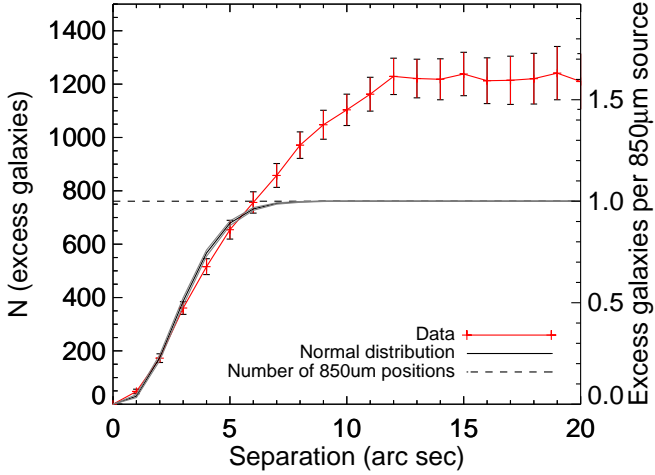}}
  \caption{The radial distribution of excess galaxies around the CLS 850\,$\mu$m positions (red points with error bars), compared with expectations based on a single counterpart per SMG assuming normally distributed positional offsets depending on the signal-to-noise ratio, following \citet{ivison07}, shown as the solid black line with grey region indicating the 16th and 84th percentiles of the expected excess based on our Monte Carlo simulations. While the left-hand vertical axis shows the total number of excess galaxies around the sample of 761 850\,$\mu$m sources in our search area, the right-hand axis has been scaled to show the average number of excess $K$-band galaxies around each source. The horizontal dashed line shows the number of 850\,$\mu$m sources in our catalogue.}
  \label{fig:multiplicity}
\end{figure}

Having ascertained that the UDS data are deep enough that almost every CLS 850\,$\mu$m source has at least one detectable (if not identifiable) and associated $K < 24.6$ galaxy, we now wish to estimate how many galaxies we find for each. We do this by calculating the number of possible counterparts within a range of search radii between 0 and 20 arc sec, and subtract off the number of unassociated background sources that we would expect based on scaling the source density in the UDS $K$ band catalogue to the relevant search area. The results of doing this are shown in figure \ref{fig:multiplicity}, in which the red points with Poisson error bars show the radial distribution of the number of excess galaxies around the 761 850\,$\mu$m sources. For comparison, we also overlay the corresponding plot based on 1,000 Monte Carlo simulations of the hypothetical scenario with one counterpart galaxy for each 850\,$\mu$m position, assuming normally-distributed positional uncertainties depending solely on the signal-to-noise ratio of each source in the CLS catalogue \citep[following the method in][]{ivison07}, added in quadrature with the 2 arc sec uncertainty resulting from pixel-quantised positions in the public CLS catalogue. We overlay the median-likelihood values (black) on the 16th and 84th percentiles of the cumulative frequency distribution (in grey) as a function of separation. We observe that the number of excess galaxies increases steadily out to around 12 arc sec, giving an average of $1.52 \pm 0.09$ excess $K < 24.6$ galaxies around each 850\,$\mu$m source. These values highlight that in many cases we are detecting an excess of nearby $K$-band galaxies rather than the sources responsible for the 850\,$\mu$m emission; this number is therefore not directly analogous to the SMG `multiplicity' discussed above, but assuming that it is not dominated by a handful of sources with many components (and our estimate of $Q_0$ derived in section \ref{subsec:q0} suggests that this is a reasonable assumption), finding $1.52 \pm 0.09$ excess galaxies per 850\,$\mu$m position is in good agreement with more direct studies of SMG multiplicity, both observationally \citep[e.g.][]{chen16}, and using simulations \citep[e.g.][]{hayward13b}.

\subsection{The properties of galaxies near 850\,$\mu$m sources}
\label{subsec:distributions}

\begin{figure}
  \centering
  \subfigure{\includegraphics[width=0.42\textwidth, trim=0.3cm 0.1cm 0.3cm 0.3cm, clip=true]{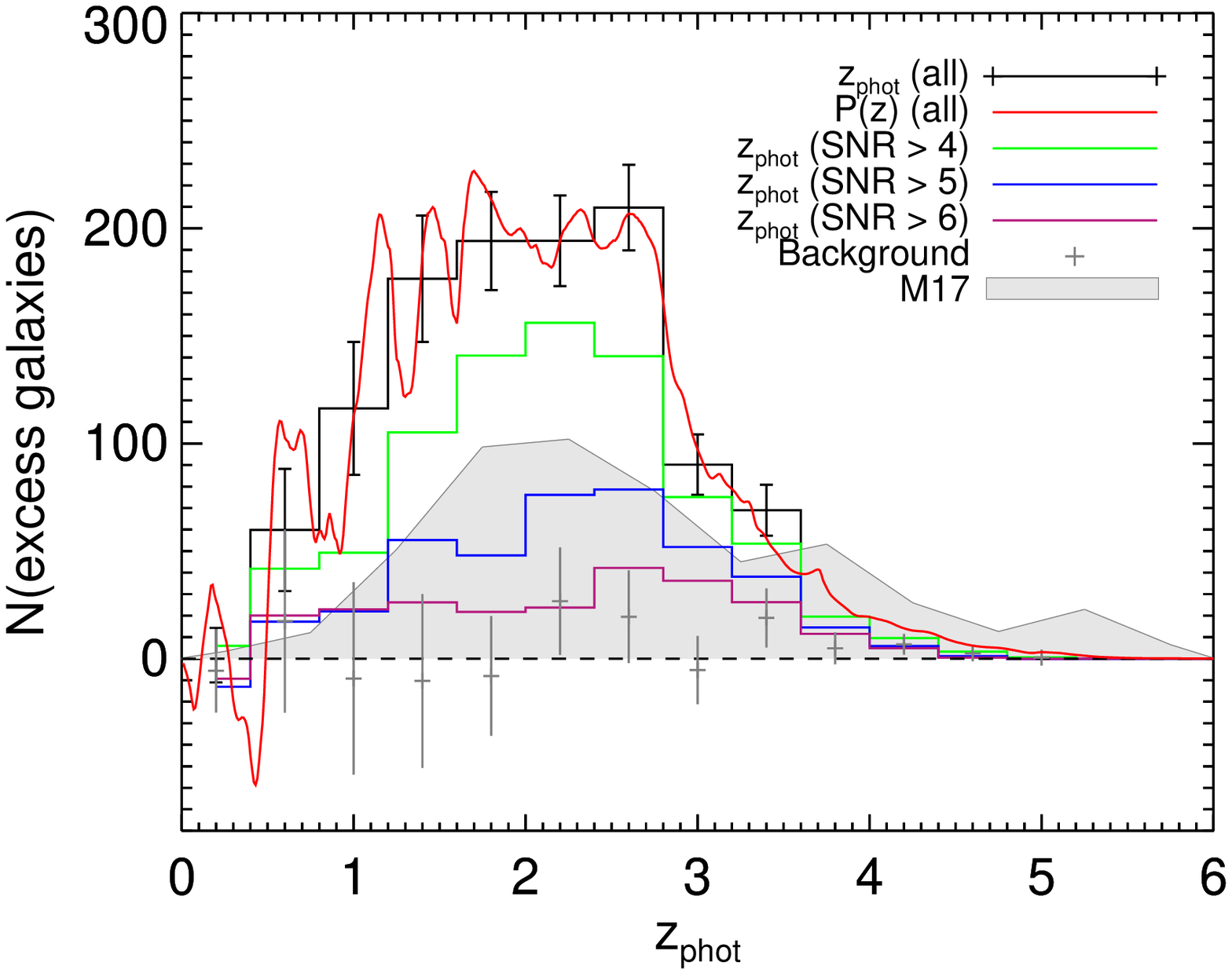}}
  \subfigure{\includegraphics[width=0.42\textwidth, trim=0.3cm 0.1cm 0.3cm 0.3cm, clip=true]{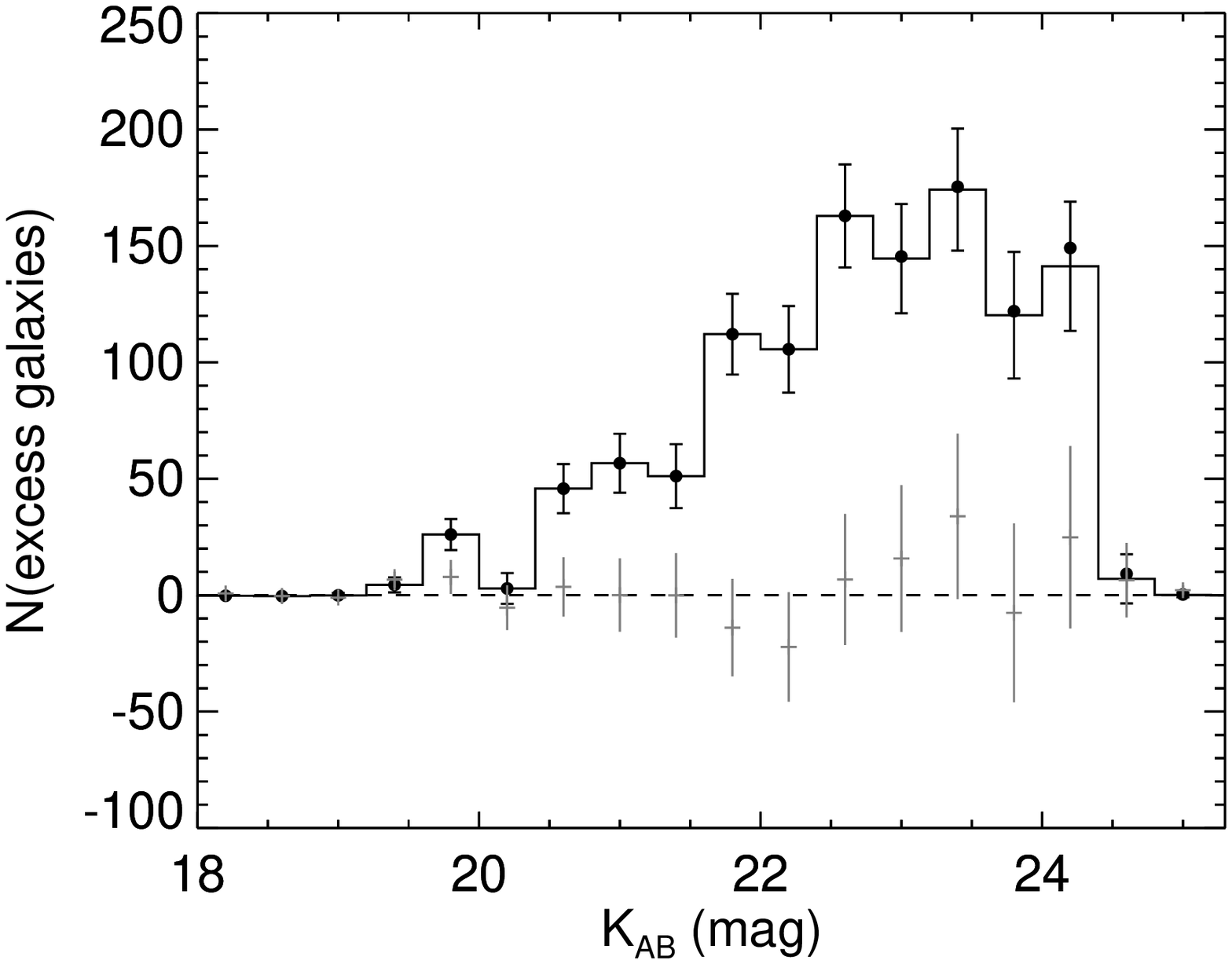}}
  \subfigure{\includegraphics[width=0.42\textwidth, trim=0.3cm 0.1cm 0.3cm 0.3cm, clip=true]{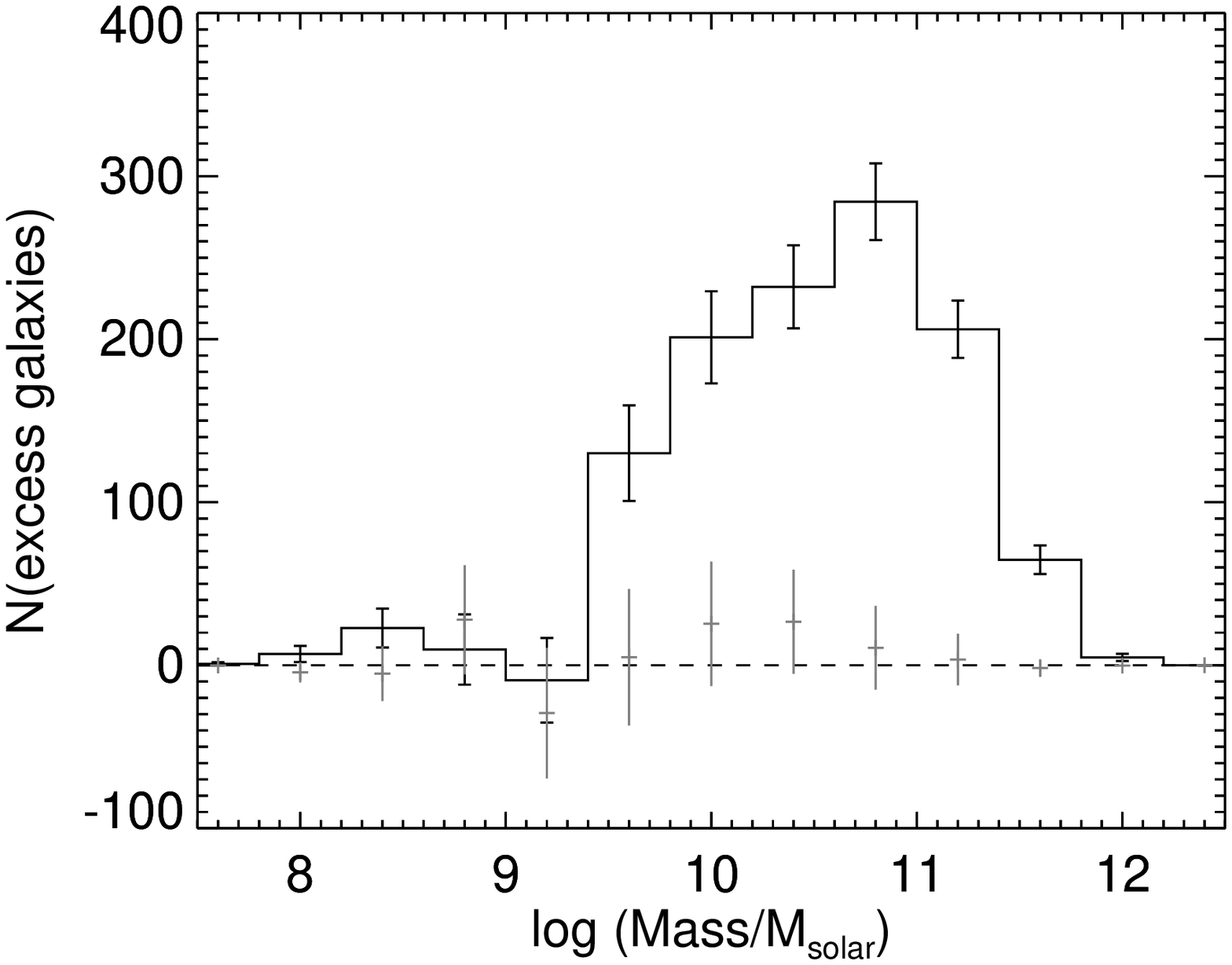}}  \caption{The properties of the galaxies within 12 arc sec of 850\,$\mu$m sources including, {\bf Top:} photometric redshift distribution (which our simulations suggest is indistinguishable from the redshift distribution of SMGs), {\bf Middle:} $K$ band magnitude distribution and {\bf Bottom:} best-fit stellar mass distribution. In each panel the grey points with error bars show the results of applying the same background subtraction method to the same number of random search radii as there are CLS 850\,$\mu$m sources in our survey area, with the error bars estimated from propagating the Poisson errors on each individual data point. In the top panel, we also overlay the photometric redshift distribution from \citet[][shown as the grey shaded region]{michalowski16}, and those that we have produced for brighter sub-sets of the SMG population, detected at $>4, >5$ \&\ $>6\sigma$, shown as green, blue and purple histograms, respectively. We also overlay the corresponding summed $P(z)$ derived based on the individual redshift probability distributions for each source (in red). In the middle panel, the black filled circles show the same data as the error bars, but corrected for residual completeness in the UDS catalogue, derived using the corrections from \citet{mortlock15}.}
  \label{fig:smg_properties}
\end{figure}

We characterise these galaxies by extending the method of section \ref{subsec:multiplicity} to produce background-subtracted distributions of their photometric redshifts, $K$ band magnitudes and stellar masses, and these are shown in figure \ref{fig:smg_properties}. To determine the extent to which the properties of these sources are representative of the SMG host galaxy population, we perform an identical analysis using realistic mock catalogues based on lightcones from the Bolshoi simulation \citep{klypin11}, containing galaxies with stellar mass $> 10^9$\,M$_\odot$ over $0.5 < z < 8.0$, which were produced by \citet{hayward13b} to study SMGs. While the stellar masses of excess galaxies within 12 arc sec of 850\,$\mu$m sources in the Bolshoi simulation (selected to the same flux criterion as in the real CLS catalogues) are significantly different from the stellar masses of the SMGs themselves, a $K-S$ test shows that we are unable to rule out the hypothesis that the {\it redshift} distributions are drawn from the same parent distribution (we perform this simulation 1,000 times and find a median $P \approx 0.3$). We therefore conclude that the photometric redshift distribution derived in this manner is representative of the full SMG population.

In the top panel of figure \ref{fig:smg_properties} we show the photometric redshift distribution of the $1154.6 \pm 67.5$ excess galaxies around all 761 $> 3.5\sigma$ CLS 850\,$\mu$m sources in our survey area as the black histogram, derived from the best-fit redshifts from the \citet{hartley13} catalogue, and overlaid (in red) with the corresponding distribution derived from the individual redshift probability distribution functions. The median redshift of the 850\,$\mu$m sources is $\langle z \rangle = 2.05 \pm 0.03$, which makes for an interesting comparison with many previous studies despite their incomplete redshift information \citep[e.g.][]{wardlow11,casey13,simpson14,chen16}, and despite the important role played by selection (which we will discuss in section \ref{subsec:brightness}). 
We find that 18.2\,per cent of the SMGs in our sample lie at $z > 3$, in good agreement with \citet{wardlow11}. We also overlay the results of performing an identical analysis on the same number of random positions as the grey error bars, which recover a redshift distribution consistent with zero excess galaxies, reassuring us that our results are not caused by some undiagnosed bias. We have overlaid our results on the redshift distribution for the $\ge 4 \sigma$ sample of SMGs selected from the CLS observations of the COSMOS and UDS fields from \citet{michalowski16}, who (as discussed in section \ref{sec:intro}) completed the redshift distribution for the $\sim$third of sources without cross-identified counterparts in their sample using long wavelength data. The \citet{michalowski16} redshift distribution (which has been arbitrarily re-scaled for clarity) has a higher median $\langle z \rangle = 2.40^{+0.10}_{-0.04}$, and this is most readily apparent in the increased high-redshift tail in the shaded grey distribution in the top panel of figure \ref{fig:smg_properties}. This apparent disagreement between the two $> 4\sigma$ source redshift distributions \citep[shaded grey from][corresponding to the green distribution from this work]{michalowski16} is of particular interest given the complementary strengths and weaknesses of the two methods. 

Though the Bolshoi simulation shows us that the mass distribution of the galaxies in the simulation around the 850\,$\mu$m source positions differs from the mass distribution of the model SMG counterparts, it is nevertheless instructive to calculate the $K$-band magnitude and stellar mass distributions of these sources.  In the middle panel of figure \ref{fig:smg_properties} we show the distribution of $K$ band magnitude for the excess galaxies, overlaid with the results of the analogous experiment using the same number of random positions (grey error bars). Interestingly, rather than increasing towards the limit of the UDS catalogue, the magnitude distribution peaks around $K = 23.4$ magnitudes ($\sim1.2$ magnitudes above the $K = 24.6$ magnitude limit) before turning down. This reinforces our result from section \ref{subsec:q0} that the UDS data are sensitive enough to detect an excess of galaxies around virtually all of the 761 850\,$\mu$m sources in the \citet{geach17} catalogue for which we have coverage. If the excess $K$-band galaxies that we detect are not the sources responsible for the 850\,$\mu$m emission, the magnitude distribution of SMG counterparts must be bimodal, as expected based on the extreme dust columns around bright SMGs found by \citet[][$A_V \approx 540^{+80}_{-40}$\,mag]{simpson15b}. That we are detecting galaxies around 850\,$\mu$m sources not necessarily responsible for the 850\,$\mu$m emission offers a possible means to reconcile our results with the fact that 5 out of the 30 $> 8$\,mJy CLS sub-millimetre sources in \citet{simpson16} do not have a $K$-band counterpart to at least one of the multiple blended components resolved by ALMA in this field.

To test for the possible influence of incompleteness in our $K$-band catalogue on our magnitude distribution, we correct the $K$-band magnitude distribution using the completeness curve for the parent UDS DR8 catalogue from \citet{mortlock15}, and the results of applying this correction are overlaid as the black filled circles, which are consistent with the original results. In the bottom panel of figure \ref{fig:smg_properties}, we show the corresponding mass distribution. We find galaxies around SMGs have an average stellar mass of $2.2 \pm 0.1 \times 10^{10}$ M$_\odot$, around a factor of two lower than the value found by \citet[][after accounting for the different initial mass function used in that work]{wardlow11}, though as noted above, that work assumed a standard $H$-band mass-to-light ratio for the stellar mass estimation, and was unable to account for the whole 850\,$\mu$m source population, while we are likely biased towards lower stellar masses by including SMG neighbours in our calculation. 

\begin{figure}
  \centering
    \subfigure{\includegraphics[width=0.48\textwidth, trim=0.6cm 0.4cm 0cm 0.4cm, clip=true]{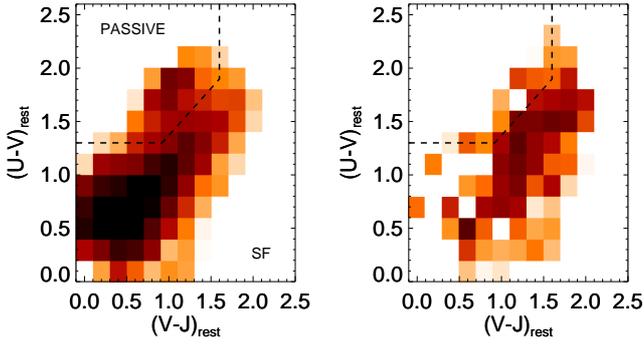}}
  \caption{{\bf Left:} The distribution of $K_{\mathrm{AB}} < 24.6$ galaxies within the $UVJ$ colour-colour space from \citet{williams09}, which is designed to separate passive and star-forming galaxies. The histograms are normalised to the area around all 761 850\,$\mu$m sources in our sample, and coloured according to the log of the bin occupancy, with values ranging from 5 to 100 galaxies per bin. The {\bf Right} panel shows the $UVJ$ colours of those $K < 24.6$ galaxies within 12 arc seconds of a CLS 850\,$\mu$m source, with the same colour scale as the left panel. 
  }
  \label{fig:uvj}
\end{figure}

Since we derived rest-frame colours for every source in the UDS catalogue, we are able to apply the same technique to determine the locations of the excess galaxies around SMGs in the $UVJ$ colour space. The results of doing so are shown in figure \ref{fig:uvj}, in which the left-hand panel shows the distribution of galaxies in the UDS catalogue, with a dashed line to highlight the division between the regions occupied by star-forming and passive galaxies at $z > 1$, taken from \citet{hartley13}. The colour of each bin is chosen to indicate the logarithm of the number of sources in that bin, ranging from 5 to 100 galaxies. In the right panel, we show the corresponding colour-colour space for galaxies within 12 arc sec of a CLS 850\,$\mu$m source, with the same colour scaling, highlighting that while the vast majority are classified as star-forming on the basis of their rest frame colours, a small subset (an excess of $92.8 \pm 23.7$, or $8.0 \pm 2.1$ per cent of the total galaxy excess) are consistent with being passive galaxies, in agreement with previous work on smaller samples of the SMGs themselves \citep{smail04}.\footnote{To convince ourselves that our background subtraction is working as expected, we inverted the data and re-plotted with the same colour scheme, revealing no significant galaxy excess.} 

These results make a particularly interesting comparison with the results of Smith et al. (in preparation), who find that the average 850\,$\mu$m flux density of $z>1$, $K$-selected and $UVJ$ star-forming galaxies increases with increasingly red $(V-J)_{\mathrm{rest}}$ colour, meaning that we expect many SMG neighbours to be sub-millimetre faint, and therefore very difficult to identify even using e.g. sensitive high resolution sub-millimetre observations with ALMA, and the same is of course true of the passive galaxies. This highlights the need for deep optical\slash near-infrared data alongside sensitive interferometry if we are to truly understand the relationship between the properties of SMG host galaxies and their environments. 

\section{Discussion}
\label{sec:discussion}

\subsection{Can we believe photometric redshifts for sources this faint?}

In deriving these results, we have of course assumed some degree of reliability in the UDS photometric redshifts, which \citet{hartley13} showed to be very precise. However, this can only be verified using spectroscopy, which is extremely challenging to obtain for large samples of faint sources such as those studied here, and the lower left panel of figure \ref{fig:smg_properties} indicates that $\sim39$ per cent of the galaxies associated with 850\,$\mu$m positions are fainter than $K = 23.0$, the limit of the UDS$z$ spectroscopic sample selection (Almaini et al. {\it in prep}). 

Meanwhile, some works in the literature \citep[e.g.][]{simpson14} have been reluctant to quote photometric redshifts for sources with formal detections (defined as having an SNR $> 3\sigma$) in fewer than four optical\slash near-infrared bands, taking the conservative view that photometric redshifts derived for such sources are unreliable. However, other works have shown the value of measurements for deriving galaxy parameters even when they are not formally significant \citep[e.g.][]{smith13,smith14}, and this is one of the reasons why photometric redshift codes such as {\sc Fast} \citep{kriek09} and {\sc LePhare} \citep{ilbert06} ideally require measured flux densities rather than magnitudes in their input catalogues (since fluxes are symmetric even for faint sources, and allow noisy negative flux estimates to be naturally included in the fitting). 

Since we have not directly identified any counterparts to the 850\,$\mu$m sources, it is difficult to directly investigate the number of photometric detections for any individual source. However, we can repeat our analysis, this time including only those sources which have at least four $> 3\sigma$ detections in the UDS catalogue, and see what difference this makes to our results. We are still able to detect at least one galaxy associated with every SMG (i.e. our estimate of $Q_0$ is still consistent with unity), and we still detect a significant excess of $1,091.8 \pm 66.8$ galaxies, as compared with $1154.6 \pm 67.5$ when the full sample is considered. This change indicates that around 5 per cent of the galaxies around SMGs have fewer than four $3\sigma$ detections, but the significance of the change is $<1\sigma$, based on propagating the Poisson error bars. The average photometric redshift of the excess galaxies around SMGs in the remaining sample is $1.99 \pm 0.03$, which is consistent with the original value of $2.05 \pm 0.03$, while the $UVJ$ passive fraction in the excess galaxies is unchanged. We therefore suggest that while this aspect may play a minor role, it does not give us any cause to mistrust our results more than we would any other study using photo-$z$; our approach benefits once more from not relying on any one single photometric redshift.

\subsection{Do the brightest SMGs differ from the rest?}
\label{subsec:brightness}

\begin{figure}
  \centering
  \subfigure{\includegraphics[width=0.49\textwidth, trim=0cm 0cm 0cm 0cm, clip=true]{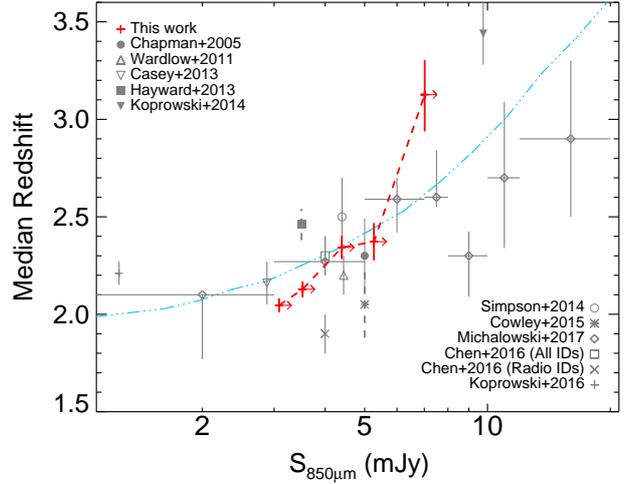}}
  \caption{Variation in the average SMG redshift as a function of 850\,$\mu$m flux density. Red error bars (connected with the dashed red line) are derived based on our sample, using the median statistics method of \citet{gott01}, while the right-facing arrows are to indicate that each data point includes all sources at higher signal to noise. Also overlaid are the average redshifts measured by selected works in the literature, detailed in the legend. Observational works are shown with solid error bars, while theoretical studies are indicated by dashed error bars. Also overlaid, as a light blue dot-dashed line, is a prediction based on the latest estimate of the far-infrared luminosity function out to $z = 5$ \citep{koprowski17}.}
  \label{fig:z_vs_snr}
\end{figure}

Several previous investigations \citep[e.g.][]{ivison02,pope05,biggs11,chen16} have found evidence for differences between the brightest and faintest SMGs in terms of their redshifts and the number of sources contributing to the single-dish 850\,$\mu$m flux density \citep[though][found no evidence for this]{wardlow11}. In figure \ref{fig:z_vs_snr}  we show how the median redshift varies as a function of the 850\,$\mu$m flux density as red error bars, ranging from the full sample, to increasingly brighter subsets (note that the left-most error bar is derived based on the whole population, with increasingly smaller subsets as we move towards brighter SNRs). Also overlaid are results from selected other observational works in the literature \citep[][]{chapman05,wardlow11,casey13,vieira13,simpson14,michalowski16,chen16}, along with their quoted error bars. We have also overlaid an estimate based on the latest measurements of the far-infrared luminosity function using SCUBA-2 and ALMA data from \citet{koprowski17}, shown as the dot-dashed light blue line) and theoretical results for bright ($> 5$\,mJy) SMGs based on the {\sc Galform} semi-analytic model presented in \citet{cowley15}, and based on the simulations from \citet{hayward13a}. 
 
We find strong evidence for a higher median redshift for increasingly bright 850\,$\mu$m sources in our sample, with error bars estimated from the redshift histograms using the median statistics method of \citet{gott01}. The average redshift ranges from $2.05 \pm 0.03$ for the full SNR $> 3.5\sigma$ sample (equivalent to a 3\,mJy-selected sample due to the highly uniform 850\,$\mu$m  map) of 761 galaxies, up to $3.13 \pm 0.18$ for the 45 sources brighter than $8\sigma$. This result becomes even clearer if we consider the values quoted by other surveys, although the scatter in those works is considerably larger due to the smaller sample sizes, with the exception of \citet{michalowski16} who calculated the average redshift in bins of 850\,$\mu$m flux density, including the ultra-faint sample from \citet{koprowski16}; these values are also overlaid in figure \ref{fig:z_vs_snr}. The brightest samples, including the $S_{\mathrm{1.1mm}} > 4.2$\,mJy SMGs from \citet{koprowski14} over the COSMOS field (corresponding to 850\,$\mu$m flux $> 9.8$\,mJy assuming a standard dust SED) with $\langle z \rangle = 3.44 \pm 0.16$, and $S_{\mathrm{1.4mm}} > 20$\,mJy SMGs (corresponding to 850\,$\mu$m fluxes $\gtsim 100$\,mJy) observed by the South Pole Telescope, which have an average redshift of 3.5 according to both \citet{vieira13} and \citet{weiss13}, further underline this trend.

\begin{figure}
  \centering
  \subfigure{\includegraphics[width=0.49\textwidth, trim=0cm 0cm 0cm 0cm, clip=true]{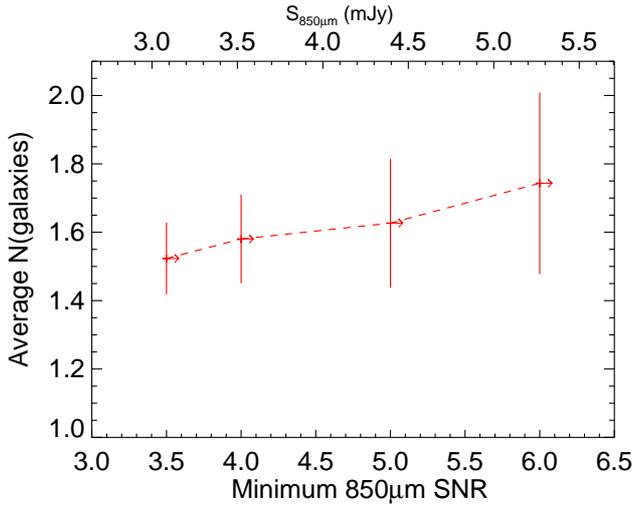}}
  \caption{The variation in the average number of $K < 24.6$ galaxies within 12 arc sec of an 850\,$\mu$m position, as a function of the 850\,$\mu$m SNR. As in figure \ref{fig:z_vs_snr}, the upper horizontal axis has been converted to 850\,$\mu$m flux density assuming a uniform uncertainty of 0.88\,mJy, the median value of sources in the CLS 850\,$\mu$m map of the UDS field.}
  \label{fig:multip_vs_snr}
\end{figure}

In figure \ref{fig:multip_vs_snr} we show a similar analysis, this time calculating the average number of galaxies around 850\,$\mu$m positions as a function of SNR\slash flux. Taking the error bars into account, the data reveal little or no evidence for an increasing number of excess $K < 24.6$ galaxies towards brighter flux densities. This is reassuring, since if brighter 850\,$\mu$m sources had a significantly larger number of excess galaxies, this might have skewed our photometric redshift distribution given that brighter sources lie at higher redshifts (figure \ref{fig:z_vs_snr}). 

It is of course possible that strongly gravitationally lensed sources within the 850\,$\mu$m selected population could artificially increase the number of excess galaxies observed around the 850\,$\mu$m positions, and previous works have shown that bright sub-millimetre flux selection can efficiently identify such systems \citep[e.g.][]{negrello10}. However, according to the latest results, lensed systems only contribute appreciably to the 850\,$\mu$m number counts above fluxes of 20\,mJy \citep{geach17}, and there are no sources this bright in our sample. Since figure \ref{fig:multip_vs_snr} also shows little evidence for an increase in the average number of excess galaxies for the brightest sources in our sample, we therefore suggest that strongly lensed galaxies do not have a major influence on our results. 

\subsection{The contribution of SMGs to cosmic star formation}
\label{subsec:csfrd}

\begin{figure}
  \centering
  \subfigure{\includegraphics[width=0.49\textwidth, trim=0cm 0cm 0cm 0cm, clip=true]{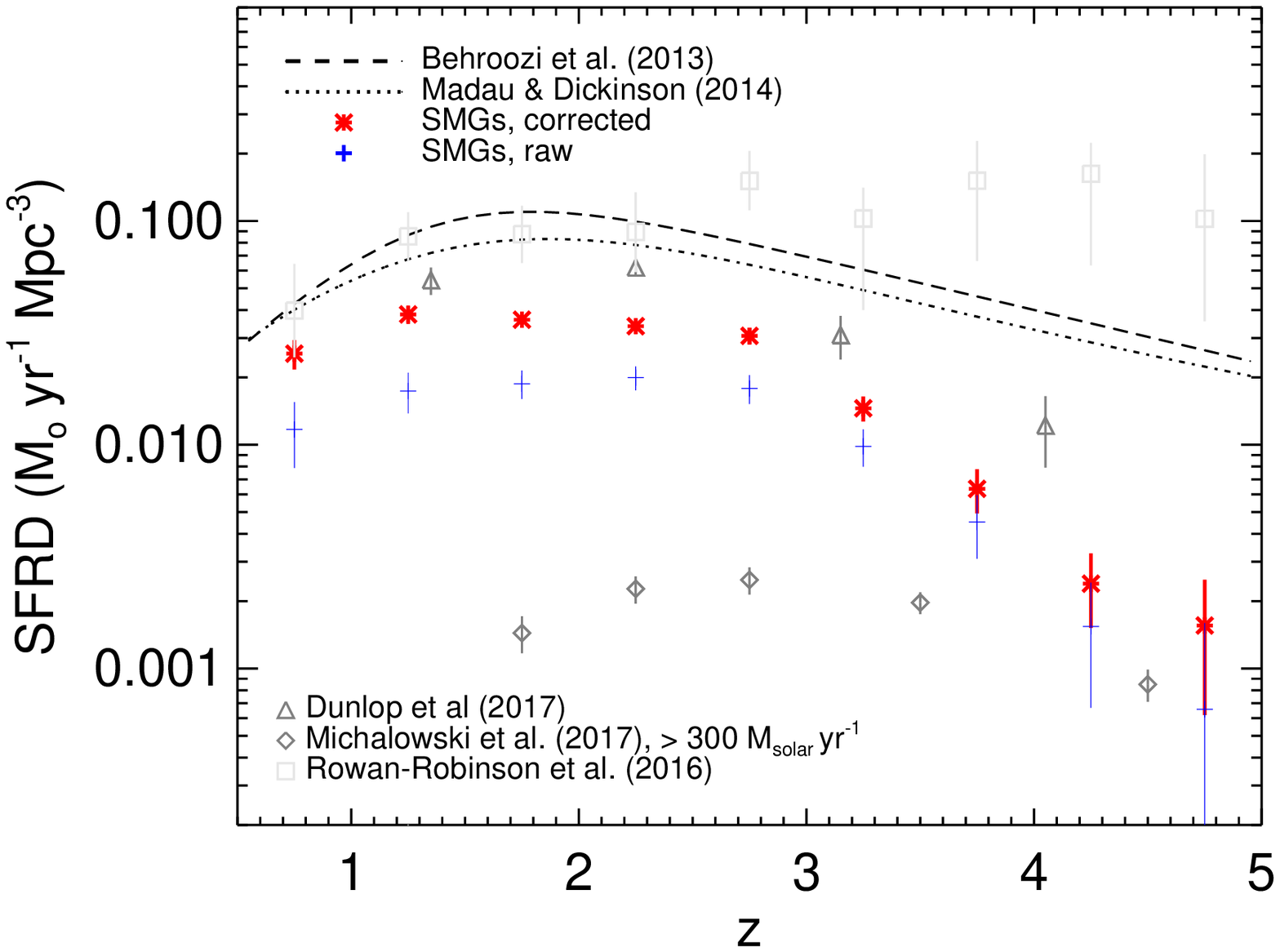}}
  \caption{The contribution of the CLS 850\,$\mu$m sources to the cosmic star formation rate density calculated by \citet{behroozi13} and \citet{madau14}, shown as the dashed and dotted lines, respectively. The raw SFRs scaled from the individual deboosted 850\,$\mu$m flux densities from \citet{geach17}, assuming an isothermal far-infrared SED with $T=30$\,K and $\beta = 1.5$ are shown as the blue crosses, while the red asterisks show the results when we fold the \citet{geach17} estimates of completenesses and false detection rates for each 850\,$\mu$m source into our calculations. The error bars are derived by measuring the 16th and 84th percentiles of the results of performing 1000 Monte Carlo realisations of sampling from the redshift distributions shown in figure \ref{fig:smg_properties}. Also overlaid are the results from the 4\,mJy SMG sample from \citet{michalowski16}, shown as grey diamonds, the obscured contribution estimated from ALMA observations of the Hubble Ultra Deep Field by \citet{dunlop16}, shown as grey triangles, and the SFRD estimated based on radiative transfer modelling of {\it Herschel} galaxies from \citet{mrr16}, shown as light-grey squares.}
  \label{fig:madau}
\end{figure}

In figure \ref{fig:madau} we bring all of this information together, in an attempt to measure the contribution of SMGs to the cosmic star formation rate density (SFRD) estimates from \citet[shown as the dashed line]{behroozi13} and \citet[dotted line]{madau14}. We do this by randomly assigning redshifts to the 850\,$\mu$m sources by sampling from the cumulative frequency distribution of photometric redshifts, using the different distributions shown in the main panel of figure \ref{fig:smg_properties} (i.e. we ensure that we encode the behaviour apparent in figure \ref{fig:z_vs_snr}).  We can then convert the deboosted 850\,$\mu$m flux density supplied in the CLS 850\,$\mu$m catalogue \citep{geach17} to an SFR assuming an isothermal dust SED model \citep[e.g.][]{hildebrand83,smith13} with $T=30$K and $\beta  = 1.5$, and a standard relationship between far-infrared luminosity and SFR from \citet{kennicutt98} adapted for our choice of initial mass function such that SFR (M$_\odot\ yr^{-1}$)$\ \approx 1.06 \times 10^{-10} (L_{\mathrm{dust}}/L_\odot)$. We estimate the star formation rate density in bins of redshift by calculating $\sum \frac{\psi_i}{V_{\mathrm{max}}}$, where $\psi_i$ represents the SFR of the $i$th source in the sample, and $V_{\mathrm{max}}$  -- the maximum co-moving volume within which that source could be detected -- is calculated by determining the redshift at which the 850\,$\mu$m source would become too faint to be included in our 3.5\,$\sigma$ catalogue, accounting for the effective area of our survey, which is 0.527\,deg$^2$. The grey points show these raw results, and we are able to correct these values to account for the completeness and fraction of false positives using the values supplied for each source in the CLS 850\,$\mu$m catalogue. The corrected values are shown as the black asterisks, with error bars assigned from repeating the redshift assignment process 1,000 times, and counting the 16th and 84th percentiles of the resulting star formation rate density in each bin. 

Though we have assumed a single isothermal dust SED template to make the conversion from 850\,$\mu$m flux density to SFR, under this assumption we find that our SMG sample with complete redshift information contributes $30.5 \pm 1.2$ per cent of the \citet{behroozi13} cosmic star formation rate density estimate between $0.5 < z < 5.0$ (where the uncertainty on the value we quote has been derived solely on the basis of Monte Carlo sampling from the photometric redshift distribution, and does not contain any contribution from the large uncertainty associated with the choice of far-infrared SED). This value is in excellent agreement with the 30 per cent estimated by \citet{barger12} based on SMGs' {\it radio} luminosities, though it is 50 per cent higher than found by \citet{michalowski10}, despite our use of a comparatively quiescent far-infrared SED relative to some other common choices (e.g. those based on prototypical local galaxies such as M82 or Arp220 from \citealt{polletta07}, or based on SMGs from \citealt{pope08} and \citealt{michalowski10}). However using any of these templates would result in the apparent contribution of star formation in SMGs exceeding the total in our lowest redshift bin at $0.5 < z < 1$. We are of course able to mitigate this behaviour by assuming a relationship between far-infrared luminosity and dust temperature \citep[as found by many other studies; e.g][]{chapman03,hwang10,smith13,symeonidis13,swinbank14}, though given that we have not identified individual counterparts to the sub-millimetre sources, it is difficult to directly address the {\it Herschel} properties of the counterparts in the current work, however this will be possible in future studies with complete x-IDs and fully deblended {\it Herschel} photometry. 

Also overlaid in figure \ref{fig:madau} are the SFRD calculated based on the obscured star formation attributed to galaxies with SFR $> 300 M_\odot$\,yr$^{-1}$ from \citet[grey diamonds]{michalowski16}, and the corresponding larger estimate of the total dust-obscured star formation based on the latest ALMA results (reconstructing the relationship between obscured star formation rate and stellar mass) using deep observations of the Hubble Ultra Deep Field by \citet[grey triangles]{dunlop16}. The comparison between the latter study and our results is of particular interest, since together the two studies imply that roughly 60 per cent of the total dust-enshrouded star formation over $1 < z < 4.5$ is captured in SMGs detected by SCUBA-2. We have also included the results derived based on a 500\,$\mu$m selected sample of sources obtained from 20\,deg$^2$ of {\it Herschel} data from \citet[as light-grey squares]{mrr16}, although as the authors note, they detect a far larger cosmic star formation rate density than other works. This discrepancy underlines the power of SCUBA-2 for high redshift studies. 

Finally, we note that despite the uncertainty over the far-IR SED, we have reached our plausible results with the largest SMG sample with complete redshift information, and having naturally accounted for the fraction of 850\,$\mu$m sources which are broken up into multiple components in high resolution sub-millimetre interferometry, even if those components are widely separated in redshift.

\section{Conclusions}
\label{sec:conclusions}

We have used 761 850\,$\mu$m sources from the SCUBA-2 CLS observations of the UKIDSS UDS field \citep{geach17}, alongside the 8th UDS data release \citep{hartley13} to statistically measure the complete redshift distribution of 850\,$\mu$m sources. Whilst our method is unable to identify the counterparts to individual SMGs, we are able to account for multiple neighbouring galaxies, and for the distribution of background sources using the rich ancillary data that are available over this field.

We find that the UDS data are sufficiently deep to detect excess galaxies around virtually all of the 761 850\,$\mu$m sources in our survey area, though we do not identify them, and they are not necessarily the same galaxies responsible for the 850\,$\mu$m emission \citep[some of which are too faint to be detected even in the UDS $K$-band data;][]{simpson16}. Simulations based on model light cones suggest that the redshift distribution of excess galaxies within 12 arc sec of an 850\,$\mu$m source is indistinguishable from that of the SMGs themselves, and we use this fact to measure their complete optical\slash near-infrared photometric redshift distribution, finding that the SMGs in our $>3.5 \sigma$ sample have an average redshift of $2.05 \pm 0.03$. We find an average of $1.52 \pm 0.09$ excess galaxies within 12 arc sec of an 850\,$\mu$m position, and use the rest-frame colours of these galaxies to show that while the majority are likely to be dusty star forming galaxies, a substantial fraction are likely to be sub-millimetre faint, passive galaxies, and therefore difficult to detect even in deep high-resolution sub-millimetre imaging e.g. with ALMA. We show that the brightest SMGs lie at higher redshifts than the rest of the SMG population, and under the assumption of a universal isothermal far infrared dust spectral energy distribution we find that SMGs contribute around 30 per cent of the cosmic star formation rate density between $0.5 < z < 5.0$.

\section*{Acknowledgments}

{The authors would like to thank the referee, Jim Dunlop, for useful suggestions that have increased the quality of this paper. We are also grateful to Will Hartley and Omar Almaini for providing the UDS catalogue used in this work, as well as to Kristen Coppin, Jim Geach and Wendy Williams for valuable discussions. 
The James Clerk Maxwell Telescope has historically been operated by the Joint Astronomy Centre on behalf of the Science and Technology Facilities Council of the United Kingdom, the National Research Council of Canada and the Netherlands Organisation for Scientific Research. The James Clerk Maxwell Telescope is operated by the East Asian Observatory on behalf of The National Astronomical Observatory of Japan, Academia Sinica Institute of Astronomy and Astrophysics, the Korea Astronomy and Space Science Institute, the National Astronomical Observatories of China and the Chinese Academy of Sciences (Grant No. XDB09000000), with additional funding support from the Science and Technology Facilities Council of the United Kingdom and participating universities in the United Kingdom and Canada. Additional funds for the construction of SCUBA-2 were provided by the Canada Foundation for Innovation. The UKIDSS project is defined in \citet{lawrence07} and uses the UKIRT Wide Field Camera \citep[WFCAM;][]{casali07}. The Flatiron Institute is supported by the Simons Foundation.}

\bibliography{cls_smg_multiplicity_refs}

\appendix

\bsp

\label{lastpage}

\end{document}